\documentclass[journal=apchd5,manuscript=article]{achemso}

\usepackage{chemformula} 
\usepackage[T1]{fontenc} 
\usepackage{siunitx}
\author{Pengji Li}
\altaffiliation{Contributed equally to this work}
\author{Chenxi Ma}
\altaffiliation{Contributed equally to this work}
\author{Jingzhong Yang}
\author{Tom N. Rakow}
\author{Xian Zheng}
\author{Eddy P. Rugeramigabo}
\affiliation[LUH]
{Institute of Solid State Physics, Leibniz University Hannover, Appelstra{\ss}e 2, 30167 Hannover, Germany}

\author{Franziska Krieg}
\affiliation[ETH]
{Institute of Inorganic Chemistry, Department of Chemistry and Applied Bioscience, ETH Zürich, 8093 Zürich, Switzerland}
\alsoaffiliation[Empa]
{Laboratory of Thin Films and Photovoltaics, Empa\textendash Swiss Federal Laboratories for Materials Science and Technology, 8600 Dübendorf, Switzerland}

\author{Gabriele Rain{\`o}}
\affiliation[ETH]
{Institute of Inorganic Chemistry, Department of Chemistry and Applied Bioscience, ETH Zürich, 8093 Zürich, Switzerland}
\alsoaffiliation[Empa]
{Laboratory of Thin Films and Photovoltaics, Empa\textendash Swiss Federal Laboratories for Materials Science and Technology, 8600 Dübendorf, Switzerland}

\author{Maksym V. Kovalenko}
\affiliation[ETH]
{Institute of Inorganic Chemistry, Department of Chemistry and Applied Bioscience, ETH Zürich, 8093 Zürich, Switzerland}
\alsoaffiliation[Empa]
{Laboratory of Thin Films and Photovoltaics, Empa\textendash Swiss Federal Laboratories for Materials Science and Technology, 8600 Dübendorf, Switzerland}

\author{Michael Zopf}
\affiliation[LUH]
{Institute of Solid State Physics, Leibniz University Hannover, Appelstra{\ss}e 2, 30167 Hannover, Germany}
\alsoaffiliation[LNQE]
{Laboratory of Nano and Quantum Engineering, Leibniz University Hannover, Schneiderberg 39, 30167 Hannover, Germany}
\email{michael.zopf@fkp.uni-hannover.de}

\author{Fei Ding}
\affiliation[LUH]
{Institute of Solid State Physics, Leibniz University Hannover, Appelstra{\ss}e 2, 30167 Hannover, Germany}
\alsoaffiliation[LNQE]
{Laboratory of Nano and Quantum Engineering, Leibniz University Hannover, Schneiderberg 39, 30167 Hannover, Germany}

\title{Deterministic Integration of CsPbBr\textsubscript{3} Quantum Dots with Plasmonic Ring Microcavities}

\abbreviations{IR,NMR,UV}

\begin{document}







\begin{abstract}
    Perovskite quantum dots hold great promise for quantum information processing as wavelength-tunable single photon sources operable over a broad temperature range. However, their deterministic integration into nanophotonic structures remains a major challenge, limited by their random spatial distribution and non-directional emission. In this work, we employ a two-step electron beam lithography process to deterministically place CsPbBr\textsubscript{3} quantum dots within the mode volume of plasmonic ring microcavities. Simulations predict strong field enhancement within the cavity, boosting photon emission rates via the Purcell effect and improving the quantum efficiency of the emitters. Experimentally, coupling ensembles of CsPbBr\textsubscript{3} quantum dots to the cavities results in a four-fold enhancement in photoluminescence intensity and a three-fold reduction in fluorescence lifetime at room temperature. Single-emitter coupling is further investigated at cryogenic temperatures, leading to a two-fold reduction in radiative lifetime. These results demonstrate a scalable approach for the integration of perovskite quantum dots into nanophotonic cavities and quantum photonic circuits.
\end{abstract}

\section{Introduction}
The rapidly advancing field of quantum photonic technologies has led to a growing focus on various quantum dot (QD) systems that possess deterministic single-photon emission property \cite{Aharonovich2016,cao2019telecom}. Metal halide perovskite QDs have recently attracted attention as promising candidates, owing to their excellent electrical and optical properties, cost-effective synthesis, wavelength-tunable emission, and wide operating range from cryogenic to room temperatures \cite{Akkerman2018,Fu2018}. Among different perovskite material systems, CsPbBr\textsubscript{3} QDs have recently demonstrated superior performance at cryogenic temperatures, exhibiting single-photon emission with an optical coherence time $T\textsubscript{2}$ of \qty{80}{\pico\second} and a radiative lifetime $T\textsubscript{1}$ of \qty{210}{\pico\second} \cite{Utzat2019}. Large CsPbBr\textsubscript{3} nanocrystals showed Hong–Ou–Mandel interference between sequentially emitted photons with a visibility of 0.56 \cite{Kaplan2023}. However, the simultaneous realization of stable and bright emission of indistinguishable photons remains a challenge. Integrating CsPbBr\textsubscript{3} QDs with optical microcavities offers a promising route to improve their performance. The radiative lifetime can be reduced by leveraging the Purcell effect, thereby improving both photon indistinguishability and collection efficiency, and unlocking the full optical potential of these emitters. Prior studies have demonstrated the coupling of colloidal QDs with optical microcavities, such as plasmonic nanoantenna \cite{Hoang2015,Hoang2016,akselrod2016efficient,nazarov2024ultrafast,iyer2021near,wang2024bright} and dielectric circular Bragg grating \cite{Jun2023,purkayastha2024purcell}, usually by drop-casting or spin-coating. However, these techniques result in colloidal QDs being distributed unpredictably with arbitrary dipole orientations. Since the emission properties of QDs are highly correlated to their exact alignment with the cavity mode, such a non-deterministic approach hinders effective coupling and makes it difficult to predict and optimize the performance. Therefore, it is crucial to develop a consistent and reliable method for deterministic perovskite QD integration.

In this study, we present a coupled system of CsPbBr\textsubscript{3} QDs with plasmonic ring microcavities (PRMs) using a two-step electron beam lithography (EBL) process. For the first time, deterministic placement of single CsPbBr\textsubscript{3} QDs inside optical microcavities has been realized. Simulations reveal that, in contrast to other nanomaterials where the primary dipole orientation significantly impacts Purcell enhancement, the total Purcell effect for CsPbBr\textsubscript{3} QDs is solely determined by their position within the cavity. Deterministic positioning of perovskite QDs is ensured by the accuracy of the EBL process, while a polymer matrix protects the QDs from moisture and air. The resulting single emitter exhibits stable, narrow-linewidth emission at cryogenic temperatures, with a two-fold reduction in photoluminescence (PL) lifetime. This approach paves the way for efficient integration of colloidal quantum emitters into on-chip photonic structures.

\section{Results and discussion}

\begin{figure}[htbp]
\centering
\includegraphics[width=0.45\textwidth]{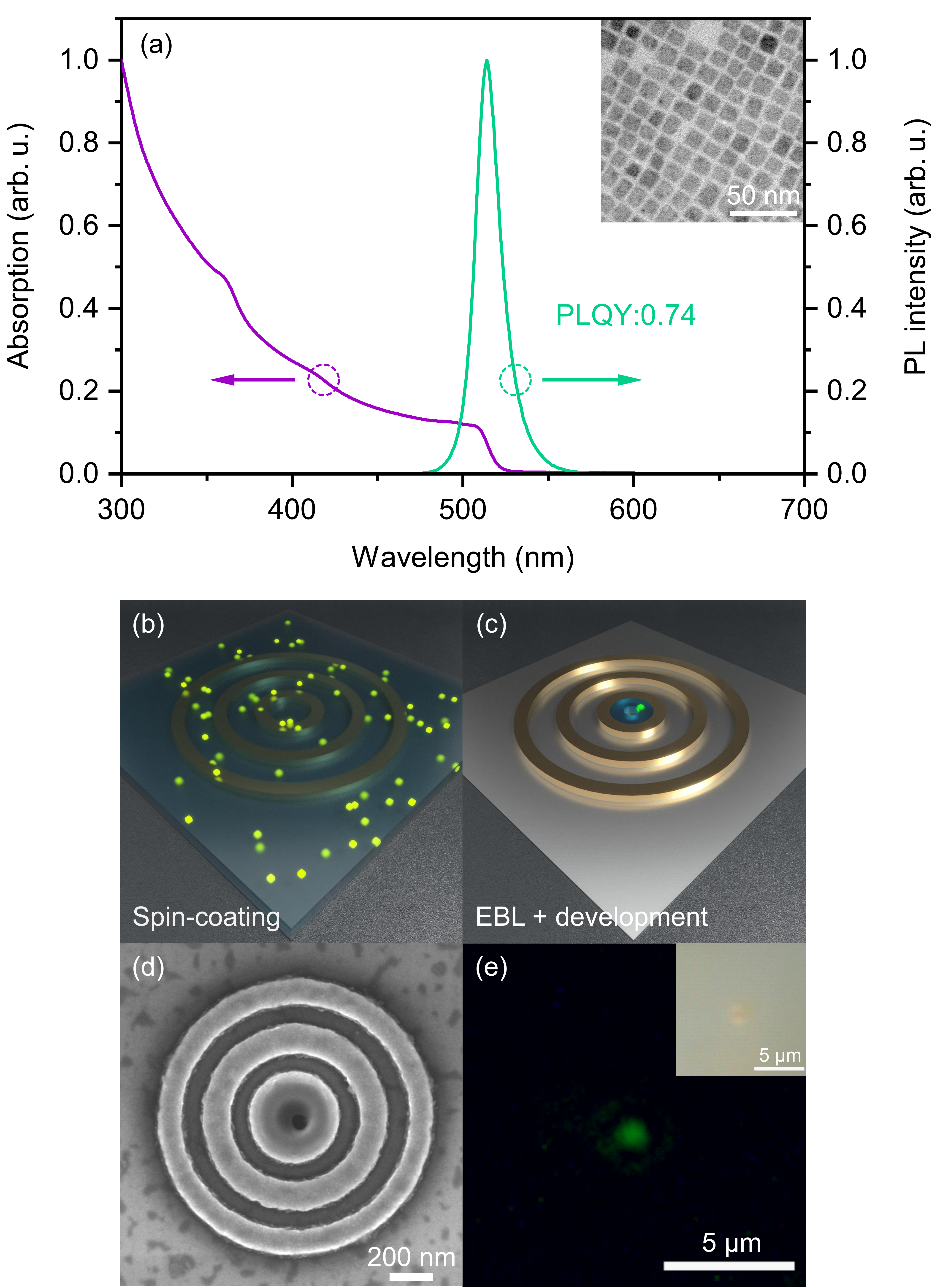}
\caption{Properties of CsPbBr\textsubscript{3} QDs and deterministic placement of single QDs. (a) PL and absorption spectra of CsPbBr\textsubscript{3} QDs (inset: TEM image). (b) Schematic of the fabrication process, starting with spin-coating a QD-containing PMMA film onto the PRM. (c) After EBL and resist development, a PMMA-QD ring remains at the inner edge of the central PRM ring as visible in the SEM image in (d). (e) PL image of a PRM showing single QD emission (inset: white-light image of the same region).}
\label{fig:1}
\end{figure}

The CsPbBr\textsubscript{3} QDs used in this experiment were synthesized by a hot-injection method and purified by centrifugation (details in Supporting Information (SI), Section A), resulting in cubic sizes from \qtyrange[range-units = single]{9}{12}{\nano\metre} as illustrated by a representative transmission electron microscopy (TEM) image shown in the inset of Fig. \ref{fig:1}(a). The optical properties of the CsPbBr\textsubscript{3} QDs in solution are demonstrated in Fig. \ref{fig:1}(a), which presents their absorption and PL spectra. This kind of perovskite QDs exhibit clear excitonic absorption and narrow PL emission at \qty{526}{\nano\metre} at room temperature, with a photoluminescence quantum yield (PLQY) of \qty{74}{\percent}\cite{krieg2019acs}.

While these optical properties make CsPbBr\textsubscript{3} QDs promising candidates for quantum light sources operating in the visible wavelength range, their deterministic integration into photonic structures remains a considerable challenge\cite{chen2022integrated,chen2021integration}. The precise spatial control of perovskite QD placement is essential to fully exploit their optical potential in nanophotonic environments. To this end, we implemented an EBL realignment process \cite{Livneh2016,Harats2017,chen2018deterministic} to fabricate QD-containing PMMA nanostructures at any desired positions with high precision. As illustrated in Fig. \ref{fig:1}(b-e), the fabrication begins with spin-coating a dilute solution of CsPbBr\textsubscript{3} QDs dispersed in PMMA resist onto the sample substrate, forming a uniform membrane encapsulating evenly distributed individual QDs, with approximately \qty{90}{\nano\metre} thickness (Fig. \ref{fig:1}(b)). Both membrane thickness and QDs density can be tuned via the dilution ratio, depending on different application purposes. Then after EBL and subsequent development, PMMA structures containing QDs remain only in the unexposed regions as shown in Fig. \ref{fig:1}(c) and SEM image in Fig. \ref{fig:1}(d), while QDs in the exposed area are washed away to avoid interference to the signal. In this experiment, ring-shaped areas near the inner edge of PRMs were strategically left unexposed to overlap with the cavity field enhancement region (elaboration will follow), covering a radial range from \qtyrange[range-units = single]{50}{120}{\nano\metre}. This approach enables the controlled placement of colloidal QDs at targeted locations. To confirm successful integration, we performed PL imaging (details in SI section A). As shown in Fig. \ref{fig:1}(e), the PL image captures emission from an individual QD accurately positioned within the PRM structure. The inset displays a corresponding white-light image of PRM, corroborating the spatial match between the emitter and the cavity.

\begin{figure}
\centering
\includegraphics[width=\textwidth]{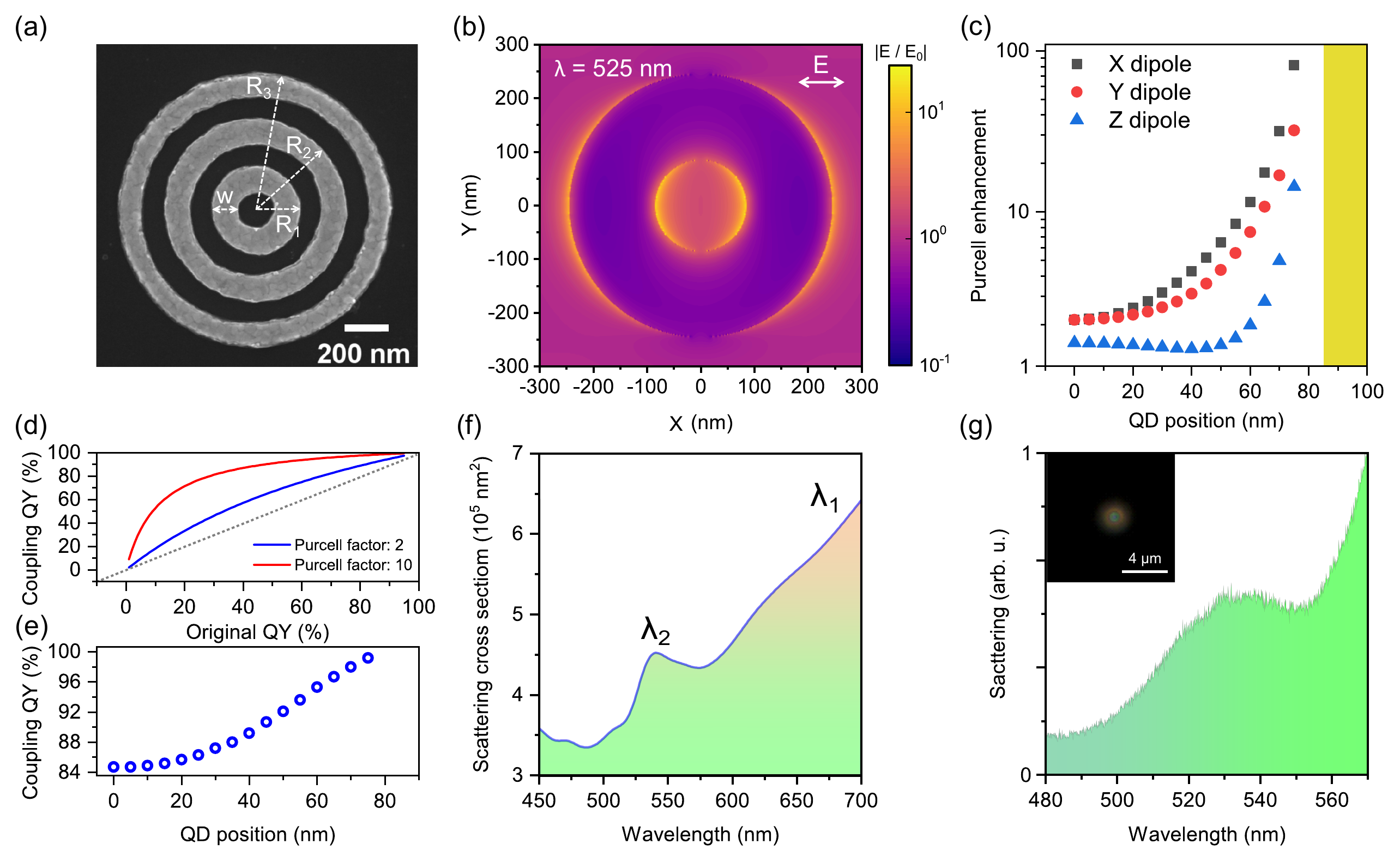}
\caption{Properties of PRMs and their coupling to QDs. (a) SEM image of a PRM. (b) Simulated electric field enhancement (normalized to the source power) in the PRM at the average emission wavelength of CsPbBr\textsubscript{3} QDs, plotted in logarithmic scale. (c) Purcell factor of the x-, y-, and z-dipole as a function of the QD position, given as the deviation from the cavity center. The yellow zone represents the position of the central ring. (d) Calculated Purcell-enhanced QY as a function of the original QY. (e) Calculated QY as a function of QD position. (f) Simulated scattering cross-section of the central ring. (g) Dark-field scattering spectrum of the PRM (inset: dark-field image of PRM).}
\label{fig:2}
\end{figure}

Plasmonic cavities are widely recognized for offering significantly stronger Purcell enhancement and extremely small mode volumes compared to their dielectric counterparts \cite{Muskens2007,baoshan2008numerical, Koenderink2017}. As a special branch, PRMs provide moderate enhancement while maintaining a relatively large mode volume\cite{cetin2012fano}, which allows for a greater tolerance of the emitter positions. They provide a controllable environment for modifying the radiative properties of fluorophores, ensuring consistent and reproducible performance in nanophotonic applications \cite{Rakovich2015}. PRMs were fabricated on silicon substrates using EBL, followed by sequential deposition of a \qty{5}{\nano\metre} titanium adhesion layer and \qty{50}{\nano\metre} gold. Fig. \ref{fig:2}(a) shows a scanning electron microscope (SEM) image of the resulting PRM structure, consisting of three concentric gold rings. The outer radii of the rings are $R_1 = 230$\,nm, $R_2 = 460$\,nm, and $R_3 = 690$\,nm, each with a nominal width of \qty{145}{\nano\metre}. Finite-difference time-domain (FDTD) simulations were conducted to numerically explore the performance of the design. To reduce the time consumption of calculation, only the central ring is analyzed in the simulation, as it primarily contributes to the results with active plasmonic response in the desired wavelength range in this experiment. A horizontally polarized plane wave ($\lambda = 525$\,nm) with average electric field amplitude $E_0$ was used as the excitation source. The electric field \textit{E} distributions from the simulation are presented on a logarithmic scale in both the top view (Fig. \ref{fig:2}(b)) and cross-sectional view (SI Fig. S1), which highlight the field confinement achieved by the design. Notably, the highest field enhancement occurs at the top inner edge of the central ring, reaching $E_{\mathrm{max}} / E_0 \approx 57.2$, which underscores the strong light-matter interaction enabled by the PRM geometry.

The Purcell effect reveals the influence of the cavity on the spontaneous emission rate of a QD by altering the local electric field distribution at its position \cite{Purcell1946,Wu2019}. Placing a QD at the field maximum of the cavity mode shortens PL lifetime, which is crucial for improving the performance of quantum light sources \cite{Lodahl2015}. In addition to spatial alignment, the actual Purcell factor is also subject to the alignment between the cavity field polarization and the dipole orientation of the QD. However, many studies on colloidal QDs coupling omit their intrinsic dipole characteristics \cite{fushman2005coupling,jiang2021single,wang2024bright,Werschler2018}. For instance, CdSe quantum rods behave as 1D dipoles \cite{xiuwen2006linear}, CdSe/CdS nanocrystals and PbS nanoplatelets may exhibit 2D dipoles characteristics \cite{Lethiec2014,li2024sub}, while PbS QDs can show 3D dipoles depending on the morphology \cite{Hu2019}. Assuming ideal alignment in both position and dipole orientation leads to overestimation of the emitter-cavity coupling strength. Therefore, realistic dipole orientations need to be considered for accurate modeling. To remedy this deficiency, we propose a model to accurately describe the interaction between a CsPbBr\textsubscript{3} QD and an optical cavity. At low temperatures, these QDs exhibit a fine-structure-split bright exciton triplet, consisting of three orthogonal, linearly polarized dipoles with high oscillation strength aligned with the crystallographic axes \cite{Becker2018,fu2017neutral,sercel2019exciton}. This splitting arises from electron–hole exchange interactions and is enabled by the orthorhombic crystal structure \cite{amara2023spectral}.

To evaluate the role of dipole orientation, we analyzed the Purcell effect for all three dipoles corresponding to the bright excitonic substates in CsPbBr\textsubscript{3} QDs in the simulation. The substrate plane was defined as the xy-plane, with the center of the PRM structure set as the origin. Accordingly, the orthogonal dipoles were assumed to align with the x-, y-, and z-axes. The dipole sources were positioned \qty{50}{\nano\metre} above the substrate to interact with the electric field enhancement region. Fig. \ref{fig:2}(c) shows the variation of the Purcell factor along the x-direction for each dipole component. The in-plane x- and y-dipoles exhibit increasing enhancement as the QD approaches the inner edge of the central ring. When far from the central ring, both dipoles show only moderate Purcell enhancement. A pronounced peak appears at a lateral offset of \qty{75}{\nano\metre}, corresponding to a \qty{10}{\nano\metre} distance from the inner edge of the central ring. At this point, the Purcell factor for the x-dipole reaches approximately 82, while the y-dipole also increases but peaks at a lower value of around 32. In contrast, the z-polarized dipole exhibits minimal enhancement throughout the entire scanning range, due to the lack of an out-of-plane component in the cavity mode. Despite this strong orientation dependence, our theoretical analysis (see SI section B) demonstrates that the total emission rate, obtained by summing over the three orthogonal dipoles, remains invariant under arbitrary rotations. This indicates that the overall Purcell enhancement is independent of dipole orientation for a fixed QD position. This orientation insensitivity is particularly advantageous given that commonly used integration methods, such as drop-casting or spin-coating, do not allow precise control over crystal orientation. Thus, our system enables consistent emission enhancement across different spatial configurations, improved robustness against fabrication imperfections, and reasonable prediction of emission rates.

In addition to modifying the emission rate, the Purcell effect also impacts the quantum yield, which is defined as:
\begin{equation}
QY=\frac{\Gamma_{\mathrm{r}}}{\Gamma_{\mathrm{r}}+\Gamma_{\mathrm{nr}}},
\end{equation}
where $\Gamma_{\mathrm{r}}$ and $\Gamma_{\mathrm{nr}}$ are radiative and non-radiative recombination rates, respectively. Upon coupling to a cavity, QY can be improved through Purcell-induced acceleration of radiative recombination \cite{Yong2017}:
\begin{equation}
QY^{\prime}=\frac{F_{\mathrm{p}} \Gamma_{\mathrm{r}}}{F_{\mathrm{p}} \Gamma_{\mathrm{r}}+\Gamma_{\mathrm{nr}}}=\frac{F_{\mathrm{p}} QY}{1-QY+F_{\mathrm{p}} QY},
\end{equation}
where $F_{\mathrm{p}}$ is the Purcell factor. This expression shows that increasing the radiative rate reduces the probability of non-radiative decay prior to photon emission, as illustrated in Fig. \ref{fig:2}(d). Using the original QY of \qty{74}{\percent} of our CsPbBr\textsubscript{3} QDs, we also calculated the position-dependent QY enhancement as a function of the QD-PRM distance. The presence of PRM significantly improves QY across a wide spatial range, facilitating more stable and efficient photon emission.

To better understand the optical resonance characteristics that contribute to Purcell enhancement, we analyzed the scattering behavior of PRM. The simulated scattering spectrum of the first ring presented in Fig. \ref{fig:2}(f) reveals two distinct resonances: a narrow resonance centered at \qty{540}{\nano\metre}, which is of interest in this study, and another much broader resonance in the near-infrared region. To experimentally verify these features, dark-field scattering measurements were performed. The measured spectrum, shown in Fig. \ref{fig:2}(e), confirms the presence of a narrow resonance peak centered around \qty{530}{\nano\metre}, closely matching the PL emission wavelength of CsPbBr\textsubscript{3} QDs. The inset displays the corresponding dark-field optical image, featuring a green central region surrounded by an orange outer ring. This chromatic contrast originates from the space-dependent scattering characteristics of PRM. The incident light angle affects the excitation efficiency of in-plane and out-of-plane plasmonic modes, and thus their relative scattering intensities, resulting in the doughnut-shaped color pattern \cite{Chow2020}.

\begin{figure}
\centering
\includegraphics[width=0.99\textwidth]{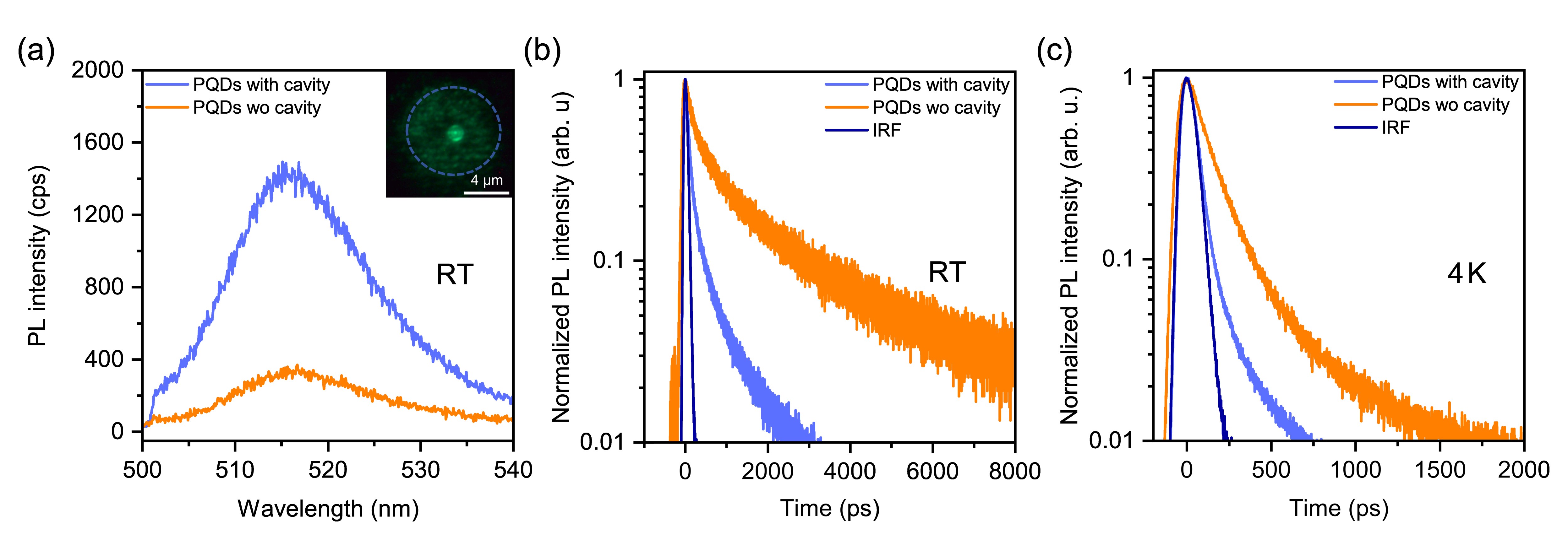}
\caption{CsPbBr\textsubscript{3} QD ensembles coupled with PRMs. (a) PL spectra at room temperature (inset: PL image showing the emission of QDs inside and outside the PRM region. The laser spot (blue dashed circle) marks the excitation area).  (b) Lifetime measurements at room temperature and (c) at \qty{4}{\K}. Perovskite QDs coupled to the PRM are shown in blue and those outside the cavity region in orange.}
\label{fig:3}
\end{figure}

Experimental validation of the coupling was performed by integrating CsPbBr\textsubscript{3} QD ensembles into the cavity and PL measurements. The QDs were incorporated using the previously described spin-coating method. A uniform PMMA film with high QD density was formed on the sample. By expanding the \qty{405}{\nano\metre} continuous-wave (CW) laser into an \qty{8}{\micro\metre}-diameter spot on the sample surface, a larger area can be excited simultaneously, enabling rapid scanning across the PRM array. The inset of Fig. \ref{fig:3}(a) presents a PL image where emission from the cavity region appears noticeably stronger than the surroundings, indicating successful coupling of QDs with the cavity. Such emitter-cavity systems are suitable for investigating cavity-modified emission behavior. We then focused the laser beam and redirected the signal into a spectrometer for $\mu$-PL measurements. Fig. \ref{fig:3}(a) shows the PL spectra from QDs coupling with a PRM and from a nearby reference region, demonstrating a four-fold enhancement in emission intensity at the PRM. We note that under weak excitation conditions, where the excitation power is well below saturation, the PL intensity can be expressed as \cite{Ge2020}:
\begin{equation}
I_{\mathrm{PL}}=N_{\mathrm{QD}} \cdot \alpha \cdot \eta_{\mathrm{exc}} \cdot QY,
\end{equation}
where $N_{\mathrm{QD}}$ is the number of excited QDs and $\alpha$ is the collection efficiency. The excitation efficiency $\eta_{\mathrm{exc}}$ is proportional to \ensuremath{{\left|d \cdot E_{\mathrm{exc}}\right|}^2}, with $d$ as the dipole moment and $E_{\mathrm{exc}}$ the excitation field amplitude. According to FDTD simulations, no substantial enhancement of the excitation field at \qty{405}{\nano\metre} is observed in the QDs region in PRM, nor is the collection efficiency improved (see SI Fig. S4 and S5). Since the original QY is already \qty{74}{\percent}, an increase of itself alone cannot account for the observed four-fold brightness enhancement. This indicates the Purcell effect not only improves QY but also accelerates the radiative decay rate and thereby enhances the brightness under CW excitation.

This led us to perform time-resolved PL measurements to confirm lifetime reduction via the Purcell effect. QD ensembles were excited using a \qty{480}{\nano\metre} pulsed laser with an \qty{82}{\mega\hertz} repetition rate, and the decay curves were fitted using a biexponential model. In Fig. \ref{fig:3}(b), QD ensembles coupled to the PRM exhibit a significant reduction in decay times at room temperature. The fast decay component is shortened by a factor of 2.8 (from \qty[separate-uncertainty-units = single]{195.4 \pm 2.4}{\pico\second} to \qty[separate-uncertainty-units = single]{69.3 \pm 0.2}{\pico\second}), with its fractional contribution increasing from 62\% to 86\%. The slow component is also shortened by a factor of 2.5 (from \qty[separate-uncertainty-units = single]{1436.6 \pm 16.2}{\pico\second} to \qty[separate-uncertainty-units = single]{573.7 \pm 3.2}{\pico\second}). At \qty{4}{\K}, as shown in Fig. \ref{fig:3}(c), the Purcell effect becomes more pronounced. The fast decay time is reduced by a factor of 3.2 (from \qty[separate-uncertainty-units = single]{126.8 \pm 0.3}{\pico\second} to \qty[separate-uncertainty-units = single]{40.4 \pm 0.4}{\pico\second}), while its contribution to the total decay process remains nearly constant at around \qty{94}{\percent}. Its slow counterpart is shortened by a factor of 2.9 (from \qty[separate-uncertainty-units = single]{621.5 \pm 6.9}{\pico\second} to \qty[separate-uncertainty-units = single]{214.0 \pm 2.9}{\pico\second}). These findings indicate that Purcell enhancement is slightly stronger at cryogenic temperatures, where radiative recombination dominates. Both fast and slow decay components are significantly affected by cavity coupling. In contrast, the long-lived tail of the PL decay (extending from \qtyrange[range-units = single]{1}{10}{\nano\second}), attributed to delayed emission caused by charging/discharging dynamics of trap states rather than direct radiative excitonic recombination \cite{becker2020unraveling}, remains unaffected. This is evidenced by the nearly identical power-law exponents observed for coupled and uncoupled samples (details in SI Section C, Fig. S6).

\begin{figure}
\centering
\includegraphics[width=\textwidth]{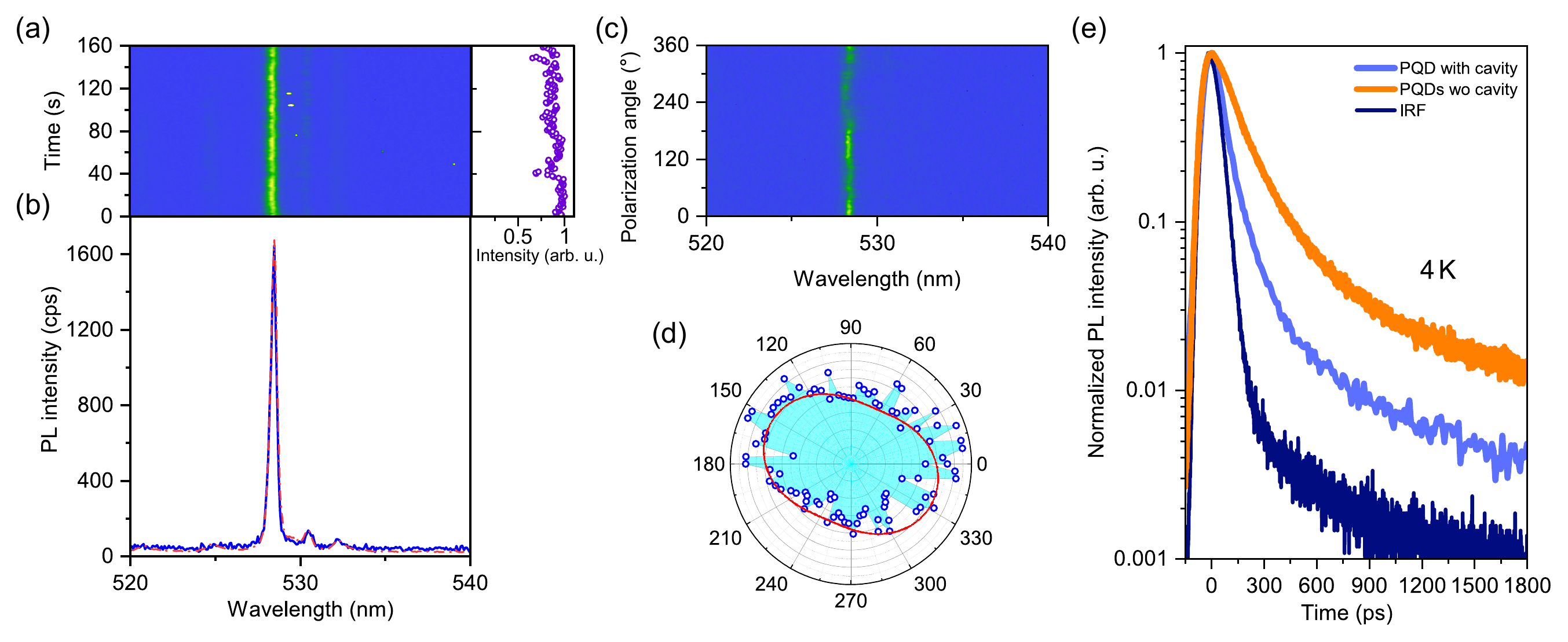}
\caption{PL characterization of a single perovskite QD coupled to a PRM at \qty{4}{\K}. (a) Time-dependent PL intensity from a coupled QD at \qty{4}{\K} and extracted PL intensity time trace on the right. (b) Corresponding PL spectrum (blue) and the average spectrum of the 161 spectra in (e) (red, dashed line). (c) Polarization-dependent PL intensity of the cavity-coupled QD at \qty{4}{\K}. The extracted PL intensity over emitted linear polarization angle is shown in (d) as a polar plot (red line: fitted curve using Eq.~\ref{eqn4}). (e) Fluorescence lifetime measurement of single QDs inside (blue) and outside (orange) the cavity at \qty{4}{\K}.}
\label{fig:4}
\end{figure}

Having demonstrated Purcell-enhanced emission in QD ensembles, we next investigated the coupling dynamics of a single QD at \qty{4}{\K}. To achieve this, we implemented the EBL realignment process as depicted in Fig. \ref{fig:1}(b-c) to precisely fabricate a PMMA/QD nanoring structure at the inner edge of the PRM. PL signals from an individual QD within the PRM structure were recorded, followed by time-trace measurements to assess emission stability, as shown in Fig. \ref{fig:4}(a). The PL intensity exhibited minimal blinking over a \qty{160}{\second} measurement period with \qty{1}{\second} exposure time per frame. The corresponding PL spectrum is presented in Fig. \ref{fig:4}(b). The blue line, representing the emission from a single \qty{1}{\second} acquisition, shows a narrow peak with a full width at half maximum of \qty{0.41}{\nano\metre}, consistent with the expected spectral linewidth of a single CsPbBr\textsubscript{3} QD at cryogenic temperatures \cite{Zhu2023}. The red dashed line denotes the normalized sum of 161 individual spectra, highlighting the spectral stability and negligible diffusion of the emission peak.

To further characterize the emission properties of the same coupled single QD, we conducted polarization-resolved PL measurements (details in SI Section A). As shown in Fig. \ref{fig:4}(c), the emission peak exhibits a clear polarization orientation, similar to the behavior of a single dipole emitter. To quantify this effect, we fitted the polarization-dependent emission using
\begin{equation}
I(\theta) = A + B \cos^2(\theta - \varphi),
\label{eqn4}
\end{equation}
where $A$ represents an isotropic background accounting for unresolved polarization components, $B$ is the amplitude of the polarized emission, and $\varphi$ denotes the principal polarization axis. The result is shown in Fig. \ref{fig:4}(d). Based on the fitted intensity values, the calculated degree of polarization $P$ is given by
\begin{equation}
P=\frac{I_{\max}-I_{\min}}{I_{\max}+I_{\min}},
\end{equation}
which yields a value of approximately \qty{16.9}{\percent}. This partial polarization indicates that the emission is not purely from a single dipole transition, but likely includes contributions from multiple bright states. Unfortunately, individual contributions from the triplet states cannot be resolved in this measurement due to the resolution limit of the spectrometer. Furthermore, under our experimental conditions with a high numerical aperture of 0.7, the tilt angle of the dipole also reduces the measured polarization contrast\cite{lethiec2014polarimetry}, imposing an additional limitation on the degree of polarization.

Finally, to evaluate the emitter-cavity coupling strength, we performed time-resolved PL measurements. As shown in Fig. \ref{fig:4}(e), the single perovskite QD coupled to the PRM exhibited a pronounced reduction in decay time compared to uncoupled QDs. The fast decay time is shortened by a factor of 2 (from \qty[separate-uncertainty-units = single]{126.8 \pm 0.3}{\pico\second} to \qty[separate-uncertainty-units = single]{62.0 \pm 0.9}{\pico\second}), with its fractional contribution slightly increasing from \qty{94}{\percent} to \qty{96}{\percent}. The slow component is also shortened from \qty[separate-uncertainty-units = single]{621.5 \pm 6.9}{\pico\second} to \qty[separate-uncertainty-units = single]{399.2 \pm 46.1}{\pico\second}. Typically in plasmonic systems, it is challenging to unambiguously attribute lifetime shortening solely to the Purcell effect, because it can also arise from enhanced non-radiative quenching when the emitter is in close proximity to metal. However, our simulations suggest that the Purcell effect in our coupling system alone is sufficient to induce a lifetime reduction far exceeding a factor of 2. These results therefore imply that the measured shortening is primarily governed by the Purcell effect rather than non-radiative Ohmic losses. Under this assumption, the moderate enhancement may be due to spatial mismatch between the QD and the cavity field maximum.

\section{Conclusion}
In summary, we have investigated the deterministic coupling of CsPbBr\textsubscript{3} QDs to plasmonic microcavities. Unlike other colloidal QDs, our calculations reveal that the total Purcell effect for CsPbBr\textsubscript{3} QDs is independent of dipole orientation but is solely governed by their position in the cavity, ensuring consistent emission enhancement regardless of the dipole tilt angle. Experimentally, ensembles of CsPbBr\textsubscript{3} QD coupled to plasmonic ring microcavities exhibited a four-fold increase in PL intensity and a three-fold reduction in PL lifetime at room temperature. To achieve precise emitter–cavity alignment, we implemented a deterministic integration strategy based on EBL, allowing the accurate positioning of single perovskite QDs at regions of high coupling strength within the cavity. The coupled system demonstrated stable, polarized emission with a two-fold reduction in lifetime at \qty{4}{\K}. This approach not only ensures reliable coupling but also offers a scalable solution for integrating perovskite QDs into photonic nanostructures. We anticipate that these results will contribute to the development of stable, Purcell-enhanced perovskite single-photon sources for future on-chip quantum technologies.


\bibliography{achemso-demo}

\end{document}